\documentstyle[12pt,aasms4]{article}
\received{1996 June 3}
\lefthead{PATTON ET AL.}
\righthead{CLOSE PAIRS OF GALAXIES IN THE CNOC1 SURVEY}

\newcommand{\arcsecpt}{\hbox to 1pt{}\rlap{\arcsec}.\hbox to 2pt{}}
\newcommand{\arcminpt}{\hbox to 1pt{}\rlap{\arcmin}.\hbox to 2pt{}}

\begin{document}

\title{Close Pairs of Field Galaxies in the CNOC1 Redshift Survey}
\author{D. R. Patton\altaffilmark{1}, C. J. Pritchet\altaffilmark{1,2}, 
H. K. C. Yee\altaffilmark{2,3}, E. Ellingson\altaffilmark{2,4}, 
and R. G. Carlberg\altaffilmark{2,3}}
\altaffiltext{1}{Department of Physics and Astronomy, University of Victoria,
PO Box 3055, Victoria, BC, V8W 3P6, Canada.}
\altaffiltext{2}{Visiting Astronomer, Canada-France-Hawaii Telescope, 
which is operated
by the National Research Council of Canada, le Centre National de Recherche
Scientifique, and the University of Hawaii.}
\altaffiltext{3}{Department of Astronomy, University of Toronto, 
60 St. George Street, Toronto, ON, 
M5S 3H8, Canada.}
\altaffiltext{4}{Center for Astrophysics and Space Astronomy, University of 
Colorado, Campus Box 389, Boulder, CO 80309-0389.}

\begin{abstract}
A redshift sample of 545 field galaxies is used to examine how the 
galaxy merger rate 
changes with redshift, and how mergers affect the 
observed properties of galaxies in close pairs.  
Close pairs are defined as those with projected separations less than 
20 $h^{-1}$ kpc.    
At a mean redshift of 0.33, it is found that 7.1 $\pm$ 1.4 \% of galaxies 
are in close 
{\it physical} pairs, compared to 4.3 $\pm$ 0.4 \% locally.  
The merger rate is estimated to change with redshift as $(1+z)^{2.8\pm0.9}$.  
These results are shown to be consistent with previous close pair studies, 
and provide one of the strongest constraints to date on the 
redshift dependence of the merger rate.   

As with earlier studies, no significant differences between the mean 
properties of paired and isolated galaxies are detected.  
However, using the subset of 
confirmed close physical pairs, those which appear to 
be undergoing interactions or mergers (based on their images) are found to
have galaxies with strong emission lines and very blue rest-frame colors.  
In addition, these pairs have low relative
velocities, increasing the likelihood that the galaxies will merge.  
We interpret this as the first clear evidence of merger-induced star formation 
occurring in field galaxies at this redshift.  

\end{abstract}


\section{INTRODUCTION} \label{intro}
Accumulating evidence indicates that galaxy populations have undergone 
significant evolution in the recent past.  
This initially became apparent with the excess number counts of faint 
blue galaxies (Tyson 1988)\markcite{TY}.  However, various redshift surveys 
(Broadhurst, Ellis, \& Shanks\markcite{BR} 1988; Colless {\it et al.} 
1990\markcite{CL90}, 
1993\markcite{CL93}) showed the redshift distribution of faint galaxies to be 
indistinguishable from no-evolution model predictions.  
The picture that is emerging from recent observations 
is one in which the number density of $L^*$ galaxies was higher at 
moderate redshift 
(Lilly {\it et al.}\markcite{CFRS6} 1995, Lin {\it et al.}\markcite{LIN} 1996).
This can be explained by the presence of star-bursting dwarf galaxies, which 
have since faded.  At present, it is unclear what is driving this process.  
One possible explanation is galaxy-galaxy merging 
(Rocca-Volmerange \& Guiderdoni\markcite{RG} 1990; 
Broadhurst, Ellis, \& Glazebrook\markcite{B92} 1992;
Carlberg \& Charlot\markcite{CC} 1992). 
Merging is expected to make galaxies brighter 
and somewhat bluer, due to enhanced star formation 
(Larson \& Tinsley 1978) \markcite{LT}, and implies 
that galaxies were less massive in the past.  However, 
the importance of merging in the 
observed evolution of galaxies remains an open question.

Galaxy merging affects a small fraction of galaxies at the present epoch.  
If the rate of merging remains constant, galaxy masses at $z$ = 1 will 
be reduced to about two thirds of their values at the present 
epoch for $\Omega$ = 1 (Carlberg 1995\markcite{RING}).  
However, there are a number of reasons 
to believe that the merger rate was significantly higher in the past.  
First, since the universe is expanding, galaxies were closer together 
in the past, so naturally merging would have been more prevalent.  
Also, recent {\it Hubble Space Telescope} ({\it HST}) imaging of galaxies at 
moderate redshift (e.g. Griffiths {\it et al.} 1994 \markcite{GR}; 
Driver, Windhorst, \& Griffiths\markcite{DR} 1995; 
Glazebrook {\it et al.}\markcite{GL} 1996) reveals a relatively high
proportion of galaxies which are morphologically anomalous, 
often exhibiting signs of interactions.  
Finally, a number of studies of close pairs of galaxies 
(e.g. Zepf \& Koo 1989\markcite{ZK}; 
Carlberg, Pritchet, \& Infante\markcite{CPI} 1994 [hereafter CPI];
Yee \& Ellingson 1995 [hereafter YE\markcite{YE}]) have 
attempted to measure the change in the merger rate with redshift; such 
studies consistently find a significant increase with redshift. 

The CNOC (Canadian Network of Observational Cosmology) cluster redshift
survey (hereafter CNOC1, see Yee, Ellingson, \& Carlberg\markcite{YEC} 
1996 [hereafter,
YEC]; and Carlberg {\it et al.}\markcite{C96} 1996) was undertaken 
primarily to study galaxy clusters.  However, 
a wealth of information on field galaxies was also compiled.  
This study utilizes the
CNOC1 catalogs to investigate the properties of close pairs of field galaxies,
and to place tighter constraints on the evolution of the merger rate.  
The availability of large numbers 
of redshifts represents an important improvement over previous studies 
of close pairs. 

In \S~\ref{obs}, the CNOC1 observations will be briefly summarized.  
Section~\ref{id} 
describes the selection procedure used to identify an unbiased sample of field 
galaxies.  The pair fractions are determined in \S~\ref{frac} and \S~\ref{ppf}, 
and the properties of galaxies 
which are paired and isolated are compared in \S~\ref{prop}.  
The redshift dependence 
of the galaxy merger rate is calculated in
\S~\ref{merge}.  The sample is divided up into redshift bins in \S~\ref{bin}
to investigate possible trends with redshift.  
We focus on the subset of confirmed physical pairs in \S~\ref{pp}.
Finally, the results and ensuing implications are discussed in \S~\ref{discuss}.
When necessary, we assume $H_0 = 100 \ km \ s^{-1} \ Mpc^{-1} \ (h=1)$ 
and $q_0$ = 0.5 throughout.

\section{OBSERVATIONS} \label{obs}
Observations for the CNOC1 redshift survey were carried out using a multi-object
spectrograph (MOS/SIS) at the Cassegrain focus of CFHT.  
The data catalogs consist of photometry (in Gunn $g$ and $r$) in the regions 
of 16 rich clusters of galaxies, lying at redshifts of 0.18 $-$ 0.55.  
Survey fields of up to 6 Mpc per cluster were covered, with the outer regions 
being dominated by the field population.  Redshifts have
been obtained for roughly half of the brighter galaxies, of which $\sim$ 50 \% 
are confirmed field galaxies; in the outer regions of the clusters, this 
fraction rises to $\sim$ 80 \%.  For a
complete description of the CNOC1 observational strategy and data reduction 
techniques,
see YEC\markcite{YEC}.
\subsection{Photometric Data} 
The basic sample consists of a catalog of photometry in Gunn $g$ and $r$, 
which is complete
to a magnitude of $r$ = 23.0 (and fainter for most frames).  
For the purposes of this study, a uniform photometric limiting magnitude 
of $r$ = 23.0
is applied.  The entire photometric survey is 100 $\%$ complete to this limit, 
and most frames are complete to $r \sim$ 23.5.
Star-galaxy classification has been performed, and is considered to 
be {\it very} reliable 
as faint as $r \sim$ 22.5.  Misclassification (mostly of galaxies as stars) 
is $\lesssim$ 10 \% at $r \sim$ 23.5 for most images, which is fainter 
than the limiting magnitude used throughout this
analysis.  This gives us a catalog with 14831 galaxies covering $\sim$ 0.66
${\rm deg}^2$.  
\subsection{Spectroscopic Data} 
The CNOC1 catalog contains redshifts for $\sim$ 2600 faint galaxies, 
of which roughly half 
are field galaxies.  Redshifts and rough spectral classification were performed 
using the cross-correlation techniques described by 
Ellingson \& Yee \markcite{EY}(1994).
The spectroscopic completeness (defined as the fraction of objects with 
redshifts) 
varies from region to region, depending on the projected number density
of cluster members and the number of MOS masks taken for a given field 
(ranging from 1$-$3).  
The cumulative spectroscopic completeness ranges from $\sim$ 50 \% at $r$ = 21.5 
to $\sim$ 17 \% at $r$ = 23.0 (here cumulative refers to all galaxies 
brighter than the given apparent magnitude).
Each redshift in the catalog has been assigned a ``significance parameter'' R, 
which is a measure of the strength of the redshift cross-correlation.  
We must consider two types of uncertainty in redshift measurements; 
identification error and velocity error.  
The former arises when the wrong peak in the cross-correlation is 
selected, and can result in catastrophic redshift errors.  
Three quarters of the redshifts have either emission-line spectra 
with R $>$ 5 or absorption-line spectra with R $>$ 4; 
such redshifts are considered to be {\it very} 
secure ($\gtrsim$ 99 \% confidence, see YEC\markcite{YEC}).  
The confidence level for the remaining objects 
drops to $\sim$ 95 \%.  As an extra precaution against identification errors, 
all redshifts
with R $<$ 3 ($\sim$ 2 \% of the sample) are excluded from the analysis.
This leaves us with redshifts for 2530 galaxies.
Once a redshift has been identified, the uncertainty in velocity can be 
estimated in 
several ways.  The most rigorous method is to use redundant observations;
this yields a typical estimated uncertainty of $\sim$ 130 km/s.
Redshift errors are discussed further in YEC\markcite{YEC}.  

\section{SAMPLE IDENTIFICATION} \label{id}
\subsection{The Primary Redshift Sample}
In order to study close 
pairs, we desire a sample of field galaxies (with redshifts) which is 
unbiased with respect to galaxy type and does not favour the identification 
of paired galaxies over isolated galaxies.  
The following sections describe the restrictions imposed to identify 
such a sample.
\subsubsection{Excluding Cluster Members}
Since we are interested in the close pair properties of {\it field} galaxies, 
we must first remove all cluster galaxies from the primary redshift sample.  
Eliminating known cluster members is a fairly straightforward procedure.  
For each cluster, a
weighted mean redshift has been identified, along with corresponding upper and 
lower redshift limits.  
These limits were derived by Carlberg {\it et al.} (1996)\markcite{C96}, using a 
manually iterated procedure which depends on each cluster's measured 
velocity dispersion.  A similar (but slightly less strict)
constraint is to uniformly exclude all galaxies with velocities within 
3000 km/s of 
each cluster's median velocity (the main results of this study are unchanged by 
imposing this criterion instead).
After this restriction, we are left with a sample of 1257 field galaxies.
\subsubsection{Excluding Field Galaxies with Potential Redshift Biases}
In the original survey, band-limiting filters were used to reduce overlap 
of spectra, and hence to maximize the number of cluster
redshifts obtained (see YEC).  As a result, the limited spectral range 
associated with each filter imposes 
a restriction on the range of redshifts for which field galaxies can be
consistently identified.  Specifically, the lower limit is set by the ability to 
detect the [\ion{O}{2}]$\lambda$3727 line within 50 $\rm \AA$ of the blue 
edge of the spectra,
while the upper limit reflects the redshift at which the 4000 $\rm \AA$ break 
is within 150 $\rm \AA$ of the red edge.  The
appropriate field redshift ranges can be found in Table 2 of YEC.  
All field galaxy
redshifts within the appropriate redshift range are expected to be 
unbiased with respect to
the redshift acquisition procedure, 
and hence are included in the primary sample.  
This leaves 797 field galaxies.
\subsubsection{Excluding the Faintest Field Galaxies} \label{self}
The CNOC1 mask design algorithm selects proportionately fewer faint galaxies,
in order to maximize the number of cluster galaxies obtained.
In addition, fainter galaxies have spectra with lower signal-to-noise ratios,
making it more difficult to extract reliable redshifts.
To compensate for this, a magnitude selection function was 
measured, which estimates the spectroscopic completeness 
(the fraction of galaxies with redshifts)
as a function of magnitude (see YEC). 
All galaxies with measured redshifts were assigned a weight, which is the
inverse of the selection function for that magnitude, 
and for that particular region.  To ensure that the primary 
field galaxy sample is unbiased, we wish to identify the range in apparent 
magnitude within which the spectroscopic 
completeness is at least 25 \%.  
Using the weights of all galaxies with redshifts, 
we determined this magnitude range for each field.  
As illustrated in Figure~\ref{figsl}, 
this amounts to identifying a faint limiting
magnitude only, since the completeness is greater than 25 \% for the
brightest galaxies.  All field galaxies which are fainter than this 
spectroscopic limiting magnitude are excluded from the primary sample.
\subsubsection{Avoiding the Clusters}
After all such restrictions have been made, 
we are left with an unbiased primary sample consisting of 572 field galaxies.  
However, there remains the danger of a given field galaxy being 
superimposed on the 
central region of a dense cluster of galaxies.  
In this case, the probability of a close companion being
an optical superposition is greatly increased.  
To avoid this situation, we wish to exclude
from the primary sample all field galaxies lying in regions which contain large 
numbers of cluster galaxies. 
This exclusion was carried out by computing the number density of 
galaxies around each field
galaxy, within a somewhat arbitrarily-defined radius of 1\arcmin.  
This radius was
chosen to be large enough to minimize the influence of 
close ($<$ 20 $h^{-1}$ kpc) companions,
while remaining small enough to detect local number density effects, 
such as the presence of
the core of a cluster.  
It is desirable to use a faint limiting magnitude (e.g. $r$ = 23.0) 
to provide good statistics.
However, to maximize sensitivity to the presence of cluster members, 
it is better to use a limiting magnitude which is closer to $M^*$.  
For the CNOC1 clusters, $M^*$ typically corresponds to $r \sim$ 20, 
and the faintest 
is $r \sim$ 21.7 (here, we have assumed $M^*_r =  -20.6$ mag 
for early-type galaxies
from local galaxy luminosity functions (LFs) [eg., 
King \& Ellis 1985\markcite{KE}]).  
At fainter magnitudes, 
the slope of the cluster galaxy LF flattens out relative to that 
of field galaxies, decreasing the fraction of cluster members.  
For this calculation only, a limiting magnitude of $r$ = 22.0 was selected.  
This includes 
an acceptable number of galaxies for computing the local number density, 
while being as close as possible to $M^*$. 

A histogram of the number density in the region of all field galaxies is 
shown in Figure~\ref{fignd}.
This plot shows the existence of a high-density tail, which is expected to 
be those field galaxies lying in the central regions of clusters.  
A density threshold of 7 galaxies per
square arcminute was selected, in order to eliminate most of the 
high-density tail 
while keeping the majority of galaxies having more typical local densities.  
In addition, all 
galaxies lying within $3\arcmin$ of a cluster center were excluded from the 
primary sample.  All plots were checked by eye to
ensure that: a) no primary galaxies remain in the central regions of clusters 
and b) no field galaxies lying 
far from the central regions are excluded from the primary sample.
Using these criteria, 27 galaxies were excluded, leaving 545
galaxies in the primary sample.  

\subsection{The Secondary Sample}
The initial task in a study of close pairs is to determine which galaxies 
have close companions.  
As with previous studies, we wish to look for companions which are comparable 
in luminosity to the primary sample, and to determine (where possible) 
which of these companions are physically associated.   
Therefore, we wish to identify a secondary
sample, consisting of all galaxies (with or without redshifts) which are 
potential companions.  In order 
to restrict the secondary sample to galaxies of similar luminosities, we need to
impose the same restrictions in apparent magnitude as were earlier applied to 
the primary sample.
Hence, we apply the spectroscopic limiting magnitudes derived in 
\S ~\ref{self} to all galaxies in the survey (with or without redshifts).  
The secondary sample is not distinct from the primary sample;
rather, the primary sample is a subset of the secondary sample.  
There are 3739 galaxies in the secondary sample, of which roughly half (1972 
galaxies) have redshifts.

\section{THE OBSERVED PAIR FRACTION} \label{frac}
In order to discuss the fraction of galaxies in close pairs, one must 
define what is meant 
by ``close pair''.  For the purposes of this study, a close pair is
defined as two galaxies with a projected separation of no more than 20 $h^{-1}$
kpc.  
At this physical separation, 
dark galaxy halos and even visible disks will interact 
significantly (Carlberg 1995\markcite{RING}).  N-body simulations indicate that 
these pairs will merge within $\sim$ 0.5 Gyr (e.g. Barnes 1988\markcite{B88}), 
as long as the galaxies have reasonably low relative velocities.
For a galaxy with a known redshift, a projected separation D translates 
into an angular separation $\theta$ using the equation 
\begin{equation} \label{eqdl}
\theta = {{D {H_0}(1+z)^2}\over{2 c \left(1+z-\sqrt{1+z}\right)}}
\end{equation}
(we assumed $\Omega_0$ = 1 for these calculations; at $z$ = 0.33, 
$\theta$ would be $\sim$ 6 \% smaller for $\Omega_0$ = 0.2).
For redshifts of 0.18 $-$ 0.67 in the primary sample, D = 20 $h^{-1}$ kpc 
corresponds to separations 
(denoted by $\theta_{20}$) of 10\arcsecpt 2 $-$ 5\arcsecpt 1, with a mean
separation criterion of $\sim$ 7\arcsec.

To calculate the observed fraction of galaxies in close pairs, 
one simply counts the number of
primary field galaxies which have companions within 20 $h^{-1}$ kpc.
Companions are selected from the secondary sample, and may or may not 
have redshifts.  For the sample of 545 primary galaxies,
73 were found to have at least one close companion.  
A slight correction was made to take into account 
the fact that a small fraction of the search area is not covered by the 
survey (due to saturated stars, column bleeding, etc.).  Using the
number of companions observed (81), 
as compared to the number we would expect to have found
if coverage had been complete (81.6), 
the expected number of galaxies with at least one companion is
then 73.6, yielding a close pair fraction of 13.5 $\pm$ 1.6 \% 
(errors assume Poisson statistics).

\subsection{Resolution Effects and the Angular Correlation Function}
Seeing for the $r$ images was typically $\sim$ 1\arcsec (FWHM).  
Hence, we are unable 
to resolve close pairs at or below this angular separation.  
To investigate empirically
what effect this may have on the calculated pair fraction, 
we have plotted the number 
density of pairs versus angular separation for the entire $r \leq$ 23.0 sample 
in Figure~\ref{figrs}a.
Three features that stand out are: (a) at large ($\gtrsim$ 7\arcsec) 
separations, the number 
density of pairs is roughly constant, as would be expected for a random
distribution of galaxies; (b) there is a significant excess 
(over {\it random}) of close pairs 
at separations of a few arcseconds; and (c) there is a clear deficit of 
close pairs at separations $\lesssim$ 1\arcsecpt 5.  

While the excess over random is significant, an excess is expected from studies
of the angular correlation function.  This function, $\omega (\theta)$, 
measures the excess number of 
galaxies (over random) found at a given angular separation $\theta$,
and can be accurately represented by a 
power law of the form $\omega (\theta) = A_{\omega} \theta ^{-\delta}$ 
(Peebles 1980) \markcite{PJE1}.
To estimate the expected $\omega (\theta)$ for our sample, we transform our 
magnitude
limit to the $F$ band, using the color transformation of Windhorst {\it et
al.} \markcite{WD}(1991) and the average color ($g-r$ = 0.82) for our sample.
Assuming the canonical value of $\delta = 0.8$ (Peebles 1980)\markcite{PJE1}, 
the amplitude for our sample is estimated to be
$A_{\omega}(1') = $ 0.056, using the observed $F$ band fits of 
Infante \& Pritchet\markcite{IP95} (1995).  For comparison, we also use their 
results for $\delta = 0.6$ (the shallowest slope found),
which gives $A_{\omega}(1') = $ 0.062.  We can then compute the expected pair 
number
density, since $\sigma_{exp}(\theta) = [1 + \omega (\theta)] \sigma_{av}$.  
Here, $\sigma_{exp}(\theta)$ 
is the expected pair number density at separation $\theta$, 
while $\sigma_{av}$ is the 
average number density of galaxies.  Note that $\sigma_{av}$ was computed using 
the observed counts at separations of $1\arcmin$, and dividing 
by 1 + $\omega(1\arcmin)$.  
That is, the observed and predicted pair densities are normalized to agree at 
$\theta = 1\arcmin$.
The predicted relations are plotted in Figure~\ref{figrs}a, and we can see that 
the observed excess is roughly consistent with what is expected  from the 
angular correlation function.  The expected $\omega
(\theta)$ is derived from a pure field sample; what effect do the 
superimposed CNOC1 cluster galaxies have on this?  
To address this issue, the analysis was repeated after excluding all 
known cluster members from the sample.  As seen in Figure~\ref{figrs}b, 
the agreement 
is similar, and indicates that the cluster galaxies do not strongly influence 
the observed relation.  

We conclude that the observed number density of pairs is 
consistent with the extrapolation of the angular correlation function to 
small scales.  We note that, while Woods, Fahlman, \& Richer 
(1995)\markcite{WD} find no excess over random 
in their photometric survey of faint pairs, 
their result appears to be consistent with the
excess found in this study.  Our Figure~\ref{figrs} shows that our results are 
clearly not consistent with no excess over random.  The deficit of close 
pairs at very small scales is confirmed, however, and is clearly due to 
resolution effects.  To compensate for this effect, a minimum search radius 
of 1$\arcsec$
was implemented when computing the observed pair fraction.  
None of the primary field galaxies 
have observed companions within this radius; hence, 
the observed pair fraction remains the same as above.  
We are unable to correct for unresolved
companions; however, by extrapolating $\omega(\theta)$ to the smallest scales, 
we would expect 
companions closer than $\sim 1\arcsec$ to contribute $\lesssim$ 5 \% to the 
observed pair fraction.  This estimate is consistent with the {\it HST} pairs 
study by Burkey {\it et al.}\markcite{BKWF} (1994; hereafter BKWF), who 
found that only 1 out of their 25 close pairs had a projected separation 
of $<$ 1\arcsec.
This effect is sufficiently small (compared to the 
uncertainty in the measured pair fraction) that we have chosen to ignore it.   

\subsection{Sampling Effects} \label{method}
When using an incomplete redshift survey to calculate the close pair fraction, 
one must sample paired galaxies fairly, or else correct for unfair sampling.
Although the CNOC1 survey is designed to compensate for pair-distribution 
selection effects (see YEC\markcite{YEC}), 
we will demonstrate a method of determining if pairs are 
{\it in fact} sampled fairly.  
In addition, this method will enable us to estimate the observed pair 
fraction in a sample which is not fairly sampled.  

We begin with an idealized sample in which all redshifts are known.  Then, by 
modelling the redshift selection procedure, one can compute what the 
observed pair properties of the sample would be.  
Suppose we start with a sample of $N$ galaxies, 
of which a fraction $x$ lie in apparent close pairs.  We model the redshift 
selection as a two-step procedure.  First, let $S$ be the probability of 
obtaining a 
redshift for an isolated galaxy or the first galaxy in a close pair.  Then, let 
$R$ be the probability of obtaining a redshift for the second galaxy in a pair, 
assuming a redshift has already be obtained for the first.  If pair selection 
is fair, $R = S$.  Now, let us also assume that, for all galaxies for which 
redshifts are obtained, 
a fraction $P$ will be members of the primary sample (i.e. lying in 
the redshift range of interest).  Knowing these properties, we can determine 
what the actual observed sample will look like.  
The observable properties we will 
use are as follows: $np_1$ is the number of primary galaxies which have 
companions with redshifts, 
$np_0$ is the number of primary galaxies which have companions without 
redshifts, $n_0$ is the number of primary galaxies which are isolated, 
and $nnz$ is the number of galaxies without redshifts.  By applying the
selection procedure, it follows that 
\begin{equation} \label{eqa}
np_1 = {N x S R P},
\end{equation}
\begin{equation} \label{eqb}
np_0 = {\frac{1}{2} N x S \left(2-R-S\right) P},
\end{equation}
\begin{equation} \label{eqc}
n_0 = {N \left(1-x\right) S P},
\end{equation}
\begin{equation} \label{eqd}
nnz = {{{1}\over{2}} N x S \left(S-R\right) + N \left(1-S\right)}.
\end{equation}
We now have four independent equations, with four input parameters ($S,R,P,x$) 
and four resultant observed quantities ($np_1,np_0,n_0,nnz$).  
For a real sample, we can retrieve the input parameters by solving this system 
of equations in terms of the observables.  
The solution is as follows:  
\begin{displaymath} 
A = {N \left(np_1+np_0+n_0\right)},
\end{displaymath}
\begin{displaymath} 
B = {n_0\left(nnz-3N\right) - 2N\left(np_1+np_0\right)},
\end{displaymath}
\begin{displaymath} 
C = {\left(N-nnz\right)\left(np_1+2np_0+2n_0\right)},
\end{displaymath}
\begin{equation} \label{eqe}
S = {{-B - \sqrt{B^2 - 4A C}}\over{2A}},
\end{equation}
\begin{equation} \label{eqf}
R = {{{np_1\left(2-S\right)}\over{np_1+2np_0}}},
\end{equation}
\begin{equation} \label{eqg}
P = {{n_0}\over{N\left(1-x\right)S}},
\end{equation}
\begin{equation} \label{eqh}
x = {{np_1+2np_0}\over{\left(2-S\right)n_0+np_1+2np_0}}.
\end{equation}

We now apply this technique to our sample, in order to determine the 
input parameters.  However, for a real sample, 
we have the complicating effect that galaxies may be found 
in multiple systems.  
That is, a primary galaxy may have more than one companion.  
For our sample, there are 65 primary galaxies with one
companion, and 8 with two companions.
We take this effect into account by using the fraction of companions 
with redshifts to calculate $np_1$ and $np_0$.  There are 49 companions 
with redshifts and 32 without; hence, $np_1$ = 44.2 and $np_0$ = 28.8, for a 
total of 73 paired primary galaxies.  The number of isolated primary galaxies 
($n_0$) is 472, giving a total of 545 primary galaxies.  
For a pure field sample, $nnz$ and $N$ are easy to determine.  With the CNOC1 
survey, we must keep in mind that primary galaxies are selected so as to avoid 
cluster centers - hence, only a subset of the area is truly surveyed for primary 
galaxies.  Instead, we compute an effective $N$ (and effective $nnz$), by 
determining the ratio of primary galaxies to total number of galaxies (and total 
number without redshifts) in the neighbourhood of each of the primary galaxies.  
An annulus with inner radius of $1\arcmin$ and outer radius of 1\arcminpt 5 
was used, which is representative of the sample as a whole.
The total number of galaxies within these annuli is 3108, of which 1329 have no
redshifts and 781 are members of the primary sample.  This leads to $N$ = 2169 
and $nnz$ = 927.  

Using equations ~\ref{eqe} $-$ ~\ref{eqh}, we find : 
$S$ = 57.1 \%, $R$ = 62.1 \%, $P$ = 43.9 \%, and $x$ = 13.1 \%.
What does this mean?  First of all, $S$ and $R$ agree closely.  If we define 
the pair sampling rate to be $R$/$S$, we find a value of 109 \%.  
This implies that galaxies in pairs are 
sampled in the same way as other galaxies, demonstrating that 
the mask design algorithm of YEC\markcite{YEC} successfully compensates for pair 
selection bias.  Also, $x$ is 
in fact the observed pair fraction for all galaxies in the sample.  
Strictly speaking, this is not 
the same as the observed pair fraction for primary galaxies, but it should
be fairly similar, since non-primary galaxies cover a similar (but larger) 
range in redshift.  
After correcting for incomplete area coverage (see \S~\ref{frac}), we 
find an observed pair fraction of 13.2 \%, in excellent agreement with the 
result found earlier in \S~\ref{frac}.   

\section{THE PHYSICAL PAIR FRACTION} \label{ppf}

\subsection{Companions With Redshifts} \label{vcut}

A total of 81 companions were found, of which 49 have redshifts.
However, a significant fraction of these companions could be optical 
superpositions.  We can use the 
available redshifts to weed out some apparent companions which lie 
at significantly different redshifts from the primary galaxies.  The aim here is
to exclude only those galaxies which cannot be physically associated with the 
primary.  Fortunately, choosing
an appropriate cutoff turns out to be quite simple.  
For all possible pairs with 2 redshifts,
2 quantities were calculated: projected separation D (in $h^{-1}$ kpc), 
and rest-frame velocity difference $\Delta V$ (in km/s). 
Projected separation is calculated with equation~\ref{eqdl}, using the angular 
separation and average redshift for the pair.
These quantities are plotted in Figure~\ref{fig2z}a, for large ranges of 
D and $\Delta V$.  Points lying in the 
lower portion of the plot represent pairs of galaxies
which are apparently close together on the sky.  
Of these, only those with relatively low velocity differences may in fact be 
physically close.  From the plot, 
there is an obvious population of close (D $<$ 20 $h^{-1}$ kpc) physical 
pairs, all of which have 
$\Delta V < $ 600 km/s; these are shown more clearly in Figure~\ref{fig2z}b.   
The remaining pairs have $\Delta V >$ 4000 km/s 
(note that velocity errors are at most 150 km/s, 
and are typically $\sim$ 130 km/s).  
Using a pair velocity criterion of 1000 km/s, we find that 22 companions 
(with redshifts) are not physically 
associated with the corresponding primary galaxy.  Hence, 55 \% of 
the companions with redshifts are deemed to be physically associated.   
If we had redshifts for all galaxies in the sample, this would allow us to 
eliminate all optical companions.  

\subsection{Companions Without Redshifts}

Even without a redshift, 
one can glean some information about a galaxy's distance from its 
color.  Specifically, 
at any given redshift, there is a limit to how red normal galaxies are
expected to be, the reddest being those of E/S0 morphological types.  
Any galaxies which are significantly redder 
than this can safely be assumed to lie at a higher redshift.  Recall
that the E/S0 envelope 
gets progressively redder with increasing redshift (cf. Koo $\&$ Kron
1992\markcite{KK}).  
This is also strikingly apparent in the color-redshift distribution of 
the field galaxies in this study (see Figure~\ref{figcc}).  
For a given field galaxy at redshift $z$, we assume that all companions 
lying more
than 0.15 magnitudes redward of the $E/S0$ sequence must lie at higher redshift.  
Note that once again we wish to eliminate only those galaxies which cannot be 
physically related, so it is best to make a conservative cut only.  
It turns out that none of the companions are eliminated using this restriction.  
However, in a larger and deeper redshift sample, this technique could be 
used to eliminate a significant number of optical companions.

We are left with 32 companions without redshifts.  
Without additional information about
these galaxies, we cannot determine which companions are 
physically associated on a galaxy-by-galaxy basis.  
However, we can correct for this effect statistically, by 
comparing the number density of the remaining companions with the background 
density of galaxies without redshifts.  
That is, we can calculate the
number of companions expected (on average) in a 
random distribution, and compare it to the
number actually found.  To estimate the number
of random companions expected, the average number density of galaxies 
(without redshifts) 
near each galaxy was computed.  Once again, we use the color cutoff.
This was done using an annulus with an outer radius of 
1\arcminpt 5 arcmin 
and an inner radius of twice $\theta_{20}$ (typically $\sim$ 15\arcsec).  
The outer radius was chosen to be large enough
to properly sample the local region around the galaxy, but small enough to 
minimize the effects
of cluster density gradients.  The inner radius was chosen to be large 
enough to avoid the 
influence of the galaxy and its close companions on the number density.
The number of random companions expected for a
given galaxy was computed by multiplying the number density by the effective 
search 
area.  Due to the resolution effects described earlier,  we use annuli
with inner radii of $1\arcsec$ and outer radii of 
$\theta_{20}$.  Any excess over random was then attributed to the presence of 
physically
associated companions.

The total number of random companions (without redshifts) expected is
16.8 $\pm$ 0.4, as compared to 32 found.  
This optical fraction (53 \%) is consistent with that 
found for companions with redshifts (45 \%), demonstrating that the statistical 
correction for optical companions is reliable (in fact, we expect the optical 
fraction
to be slightly higher for companions without redshifts, since galaxies 
outside the 
primary redshift limits are less likely to have redshifts).  
It follows that the overall fraction of
companions (with or without redshifts) which are physical is 52.2 $\pm$ 8.6 \%.  
Therefore, the fraction of galaxies which are
in close {\it physical} pairs is 7.1 $\pm$ 1.4 \%.  

\section{PROPERTIES OF GALAXIES IN CLOSE PAIRS} \label{prop}
In order to investigate the properties of galaxies in close pairs, 
we wish to compare the ``paired'' 
galaxies (those in the primary sample with companions of comparable luminosity) 
with galaxies which are not in pairs.  
For this part of the analysis, 
a ``paired'' galaxy is defined as one which has at least 
one companion which may be physically associated.  
Hence, known optical companions will be discarded.  
In identifying an isolated sample, 
we note that some galaxies have companions which are 
too faint to be included in the secondary sample.  
If we include these galaxies in the 
sample of galaxies not in pairs, 
we may be smearing out any observed property differences.  
Instead, we define the ``isolated'' sample to be those galaxies in the 
primary sample which have no companions 
brighter than the limiting magnitude of the entire sample ($r$ = 23.0).  
Galaxies with faint companions only (hereafter designated ``paired [faint]'') 
will be included in the ensuing analysis for 
completeness, but we will focus on comparing the paired and isolated samples.
The mean properties are summarized in
Table 1 (unless otherwise stated, errors quoted are errors in the mean).  
The results of various Kolmogorov-Smirnov (K-S) tests are presented in Table 2.
The K-S test statistic given indicates the significance level for the hypothesis 
that two sets of data are drawn from the same distribution.  A small 
significance level indicates that two distributions have significantly different 
cumulative distribution functions.

\subsection{Redshifts}
Of primary interest is the redshift distribution of the two samples.  
YE \markcite{YE} found that the
average redshifts of their paired and isolated samples were identical to 
within ${\Delta
z\over z} \lesssim 0.1$, implying that both populations have similar 
characteristic luminosities.   
With about 5 times as many redshifts, we can place even tighter constraints
on this relation.  The average redshift for our paired sample 
is statistically equivalent to that of the isolated sample.
We conclude that the average redshifts of the two samples are identical to 
within ${\Delta
z\over z} \lesssim 0.05$.  A K-S test and the visual appearance of the 
redshift histograms (see Figure~\ref{figzh}) 
are consistent with this conclusion.

\subsection{Galaxy Classification} \label{scl}
Each field galaxy was classified using its $g-r$ color and redshift.  The
basic method was to use a fit to the color $k$-corrections of YEC\markcite{YEC}, 
who convolved typical
galaxy spectra from Coleman, Wu, \& Weedman\markcite{CWW} 
(1980; hereafter CWW) with the filters used in the observations.  
The color-redshift relations corresponding to elliptical, Sbc, Scd, and Im
galaxies are plotted in Figure~\ref{figmc}.  
Each primary galaxy's location on this plot was used to
estimate the most appropriate type classification by 
using the sequence with the closest color at the galaxy's redshift.  
The relative proportions of early and late-type galaxies were determined for 
both the paired and isolated samples.  
The fraction of late-type (Scd+Im) galaxies is slightly  
lower for the paired sample, 
but the difference is not statistically significant.  
The corresponding K-S test result is consistent with this conclusion. 

We can perform an independent check on this result, 
using the spectral classification given
in the CNOC1 catalogs.  
Each galaxy has been assigned a spectral type based on the template
giving the highest redshift cross-correlation value.  
Galaxies are identified as either
early-type, Sbc, or emission-line galaxies.  
This determination is independent of color, and 
hence provides a good comparison with the classifications determined above.
We find that the fraction of emission-line galaxies is slightly 
lower for the paired sample, but again the difference is not significant, 
as demonstrated by the K-S test.
Hence, neither spectral nor color classification reveals any 
measurable differences  between paired and isolated galaxies. 

\subsection{Colors}
Since we might expect galaxies in close pairs to have different colors than 
other galaxies, 
we compare the observed color distributions of the two sets of data.
The mean $g-r$ colors were found to agree quite closely, 
and a K-S test confirms this.
This result has also been found by CPI\markcite{CPI} and YE\markcite{YE}, 
among others.  It is more instructive to compare the {\it rest-frame} 
colors of the two samples.  Using the spectral classifications from above, each
galaxy's color was $k$-corrected, by interpolating in color between the 
closest typical galaxy sequences. 
We find that the mean rest-frame colors are nearly identical, 
with a very slight trend toward paired galaxies being redder.
No apparent difference in the distribution of rest-frame 
colors is apparent from the histograms (Figure~\ref{fig0h}) or the K-S test.  
Therefore, we find
no measurable differences between the $k$-corrected colors of paired and 
isolated galaxies.  
Mean rest-frame color differences of $\mid \Delta (g-r) \mid \gtrsim 0.1$ 
magnitudes can be ruled out at the 3$\sigma$ level.

\subsection{Luminosities}
The apparent magnitude distributions of the two samples are also very similar.
In order to better compare luminosities, we calculate the absolute magnitude 
for each galaxy.  
Since we are only looking at relative luminosities, the results will not be 
sensitive
to the choice of cosmological parameters.  
The $k$-correction in $r$ for each galaxy 
was extracted from the CNOC1 database.  
These $k$-corrections were derived using the data 
of Sebok (1986)\markcite{SBK} for Gunn $r$, 
and depend only on the galaxy's redshift 
and color.  The average $k$-correction for the 
paired sample (0.27 mag) is identical to that for the isolated sample.

The resulting mean absolute magnitude of paired galaxies 
was found to be nearly identical to that of the isolated sample.  
This is confirmed with a K-S test (see Figure~\ref{figmh} also).  
A similar finding was made by YE\markcite{YE}.  
However, there is an indication that galaxies 
in the paired (faint) sample are significantly more luminous than galaxies 
with no companions (isolated).  This result is confirmed with a K-S test.  
The difference can be attributed to an inherent bias in selecting 
galaxies for the paired (faint) sample.  
Primary galaxies which are brighter in apparent magnitude will 
be more likely to have faint physical companions detected than will 
apparently fainter galaxies.  This selection
effect only comes into play when the apparent magnitude limit for companions is
extended deeper than that for the primary sample, and hence does not affect the 
paired sample.  

\subsection{[OII] Equivalent Widths}
One of the best indicators of current star formation is the presence of 
spectral emission lines such as [\ion{O}{2}]$\lambda$3727.  
Using a local sample of galaxies, Liu and
Kennicutt\markcite{LK} (1995) found that the mean [\ion{O}{2}] 
equivalent width was 
19 $\rm \AA$ for merging galaxies, as compared to 11 $\rm \AA$ for 
the complete sample.  We used 
the procedure outlined in Abraham {\it et al.} (1996\markcite{A2390}) 
to measure the [\ion{O}{2}] rest-frame equivalent widths of all the CNOC1 
galaxies.  We were able to determine equivalent widths 
for 81 \% of the primary sample.
Typical errors are estimated to be $\lesssim$ 20 \%.
The mean [\ion{O}{2}] equivalent width is found to be the same for paired 
and isolated
galaxies.  K-S tests and the visual appearance of the histograms 
(Figure~\ref{figeh}) 
confirm this finding.  CPI\markcite{CPI} found a similar result.  

\subsection{Properties within Sub-samples}
The paired galaxy sample can be divided into two subsets : 
those which have at least one
confirmed companion (``paired [phys]'', based on velocity criterion 
of \S~\ref{vcut}), and those for which the companions have photometric 
information only (``paired [no z]'').  
All pairs in the former sample are physical (although not necessarily 
merging) and therefore would be 
expected to exhibit greater mean property differences from the isolated 
sample if merging affects galaxy properties significantly.  
The observed properties of these two samples (see Table 1)
are found to be statistically equivalent to the isolated sample, strengthening 
the earlier conclusions.
One might also ask how the properties of the companions themselves 
compare to the paired and isolated galaxies in the primary sample.  
Due to the nature of our sample selection (see Section~\ref{id}), 
most of the companions which have redshifts are also part of 
the paired primary sample; therefore the mean properties are very similar.  
The mean apparent magnitude and color of the companions without redshifts are 
also found to be statistically 
equivalent to that of the isolated galaxies in the primary sample.  
To summarize, there are no observed differences in the overall properties of 
galaxies which have 
close companions of comparable luminosity and galaxies with no companions 
brighter than $r$ = 23.0.  

\section{THE GALAXY MERGER RATE} \label{merge}

As mentioned earlier, the average redshift of the sample is 0.33, 
which gives us some
leverage in investigating the redshift dependence of the merger rate.  
For comparison, we
will use a low redshift sample which is derived from the UGC catalog 
(Nilson 1973\markcite{UGC}).  
We select all galaxies with measured diameters of at least $1\arcmin$ 
in both $B$ and
$R$, to a (statistically complete) limiting magnitude of $B$ = 14.5.  
Within this sample, the average redshift (for galaxies with measured 
redshifts) is 0.0076.
At this distance, a physical separation of 20 $h^{-1}$ kpc corresponds to 
an angular separation of 184\arcsec.  
We find that 130 out of 
3058 galaxies have companions within this radius, yielding
a close pair fraction of 4.3 $\pm$ 0.4 \%.  
Since optical companions are negligible at 
these bright magnitudes (see CPI\markcite{CPI}), we take this to be the fraction 
of galaxies in close physical pairs.  
We note, however, that this estimate is approximate in nature.  
While the internal
error in the local pair fraction is small, the systematic errors may be 
fairly large.  A more complete redshift sample with well-defined 
selection effects is needed to improve this estimate.  

Before comparing the two samples, we must ensure that the pair 
fraction does not depend
strongly on the luminosities of the galaxies involved.  
For the nearby sample, $B$ = 14.5 corresponds to an average absolute 
limiting magnitude
of $M_B = -17.3$ (based on the average redshift).  
The CNOC1 primary field sample has an 
average apparent limiting magnitude of $r$ = 21.3, which 
corresponds to B = 23.1 using 
the Gunn color transformation of Windhorst {\it et al.}\markcite{WD}(1991) and 
the average color ($g-r$ = 0.79) of the primary galaxies.  
The average $B$ $k$-correction ($k_B$ = 1.0) was estimated using the 
mean redshift ($z$=0.33), the 
color classification fractions determined in Section~\ref{scl}, 
and the $k$-corrections of Frei \& Gunn (1994)\markcite{FG}.
This results in an average absolute limiting magnitude of $M_B = -18.1$, which
is roughly one magnitude brighter than the low-redshift sample.
It has been shown by YE \markcite{YE} that
the UGC pair fraction remains basically unchanged over a range of $\sim$ 2 
mag brighter than B = 14.5; 
hence, we conclude that the two samples are of comparable luminosity, 
and the pair 
fractions derived should not be sensitive to the particular 
magnitude limits used.  

We need to determine the fraction of galaxies in each sample which are 
likely to merge.
While those in close physical pairs are likely candidates, some may be 
moving at such high
relative velocities that they are unlikely to merge.  Using dynamical arguments, 
CPI\markcite{CPI} argue that, at the present epoch, 
pairs with relative velocities 
less than 350 km/s are likely to merge.  They estimate this fraction to 
be $\sim$ 50 \%.  
For a pairwise velocity evolving as $(1 + z)^{-1}$ (see CPI\markcite{CPI}), 
this increases to a fraction of 66 \%
at $z$ = 0.33.  There are 13 close physical pairs of primary galaxies in our 
distant sample for which redshifts 
are available for both members.  
In this sample, 11 out of 13 pairs (85 $\pm$ 26 \%) 
have $\Delta V < $350 km/s,
which is consistent with the assumed pairwise velocity.  
Using the close pair fraction determined earlier in this section, we
conclude that the local merging fraction is 2.1 $\pm$ 0.2 \%.  
It follows that, with 7.1 $\pm 1.4 \%$ of our distant sample in close 
physical pairs, 
the merging fraction at $z$ = 0.33 is $4.7 \pm 0.9 \%$.  
If we assume the functional form of $(1+z)^m$ for the increase of
the merger rate with redshift (Carlberg 1990\markcite{C90}), 
we find m = 2.8 $\pm 0.9$.  

This value is consistent with the merger rate of CPI\markcite{CPI}, who found 
m = 3.4 $\pm$ 1.0 using the same method (and the same low redshift sample).  
We note, however, that they overestimated the local pair fraction by 
underestimating its mean redshift.  
After correcting for this, we find m = 4.0 $\pm$ 1.0 for their sample.  
Our merger rate is significantly lower than that obtained 
using the physical pair fraction found by YE\markcite{YE}, which translates into 
m $\sim$ 5.0 $\pm$ 1.5 using our parameterization of the merger rate with 
redshift.
Their sample is the largest pair redshift survey next to ours; hence, it is 
important to find out what is causing this discrepancy.
One major difficulty with the YE\markcite{YE} redshift sample is that pairs 
are known to be under-selected.  
This effect was accounted for by randomly discarding redshifts 
in each two redshift pair, and then correcting for this.  
Their result was tested using the method outlined in \S~\ref{method}.  
For the observed
quantities, we use their original sample (before redshifts were discarded).  
Using $np_1$ = 7.2, $np_0$ = 9.8, $n_0$ = 90, $nnz$ = 200, and $N$ = 376, 
we find 
S = 47 \%, R = 41 \%, P = 61 \%, and x = 16.3 \%.  The pair sampling (R/S) is 
87 \%.  YE\markcite{YE} estimate that $\sim$ 55 \% of their observed pairs are 
physical.  This leads to a physical pair fraction of 8.9 $\pm$ 3.9 \%, which is 
considerably lower than the 15.5 $\pm$ 6.6 \% reported in their study.  
This revised pair fraction leads to a
merger rate of m = 3.4 $\pm$ 1.9, which is consistent with our result.  

Using {\it HST} images, BKWF\markcite{BKWF} derived m $\sim$ 2.5 $\pm$ 0.5, 
which would appear to agree with our merger rate.  
However, they parameterized the evolution of the 
merger rate in a different way (by taking the merger rate to be 
the $z$-derivative of the pair counts).  
If we use their pair fraction with our method, we find m = 4.5 $\pm$ 0.5, which 
is significantly higher than our result.  
However, YE\markcite{YE} point out that
the pair fraction of 34 $\pm$ 9 \% used by BKWF\markcite{BKWF} 
is approximately twice 
as large as the {\it physical} pair fraction (due to optical pairs).  
After taking this into
account, and using our parameterization of the merger rate evolution, we find 
m $\sim$ 2.9 for the BKWF\markcite{BKWF} sample, in agreement with our result. 

\section{REDSHIFT-BINNED SAMPLES} \label{bin}
An additional test of the validity of the preceding results is to divide the 
sample into redshift bins.  
The idea here is to look for redshift dependence in various properties, 
and also to have independent samples to look for common trends.  
We wish to have redshift bins which are 
small enough to avoid properties varying with redshift and large enough to 
ensure good statistics.  Hence, we designate the 
following two samples : 0.18 $< z <$ 0.30 and 
0.30 $< z <$ 0.45.  This spans all primary field galaxies with $z \leq$ 0.45, 
and avoids the sparsely populated high redshift regime 
(see Figure~\ref{figzh}).  Both samples have $\sim$ 240 field galaxies,
allowing for good statistics.  Note that both bins cover 
lookback time intervals of $\sim$ 0.68 
$h^{-1}$ Gyr (assuming $\Omega_0$ = 1), which each correspond 
to $\sim$ 10 \% of the age of the universe for that model.  
The results of the following comparisons 
of various properties are summarized in Tables 2 and 3.

\subsection{Galaxy Properties}
For each sample, there is no measurable difference between the observed or 
rest-frame colors of paired and isolated galaxies.  
This is consistent with the findings of
YE\markcite{YE}, who used similar redshift bins. 
In the low redshift
bin, paired galaxies are slightly more luminous on average, 
but the difference is at the one sigma level.  
The mean absolute magnitudes are statistically equivalent in the 
high redshift bin.  
No significant differences were found in the classifications of paired and 
isolated galaxies in either redshift bin  
(note that very high K-S test significance levels are the result of the discrete 
classification systems).  YE\markcite{YE}, on the other hand, 
found a larger emission line fraction in their high-redshift
pair sample, but the difference was at the one sigma level.  

\subsection{Pair Fractions}
As with the entire sample, we wish to investigate the pair fraction.  
However, by using the
same apparent magnitude criteria as before, 
we encounter the problem that the low redshift
sample will contain intrinsically fainter companions than the 
high redshift sample, rendering
a direct comparison invalid.  In order to compare the two redshift bins, 
we instead impose 
different limiting magnitudes, to sample galaxies of comparable luminosity.  
For this comparison, 
we relax the constraint that the secondary samples be drawn from populations of
similar luminosity as the primary sample.  
For the high redshift bin, we impose the faintest
magnitude limit possible - $r$ = 23.0.  
We can estimate the absolute magnitude of the faintest
companions detectable in this bin by using $z$ = 0.45 and the average 
$k$-correction of 0.27 mag, yielding $M_r = -18.1$.  For the low redshift bin, 
the average $k$-correction is 0.19 mag.  
In order that the faintest companions in the low redshift bin also
have $M_r = -18.1$, it follows that we must impose an apparent magnitude limit 
of $r$ = 22.0. 
Since all primary galaxies are brighter than this, we can do this without
reducing the size of the primary sample.  
 
Using the two defined samples, the pair fractions were determined.  
The fraction of galaxies
in close physical pairs for the low redshift bin is 10.2 $\pm 3.2  \%$, 
as compared to 
11.3 $\pm 3.5 \%$ for the high redshift bin.  
The pair fractions for the
redshift binned samples are larger than the fraction derived earlier for the 
entire sample, 
but this is to be expected since fainter companions have been included.
A more suitable comparison is the merging
fraction, found to be 6.3 $\pm 2.0 \%$ for the low redshift bin, 
as opposed to 7.6 $\pm 
2.4 \%$ for the high redshift bin.  
In principle, these two independent quantities can be used to measure the merger 
rate evolution; however, the small numbers of galaxies makes this measurement 
(m = 2.1 $\pm$ 5.2)
nearly meaningless for our sample.  
However, larger redshift surveys will be able to 
place tighter constraints on the redshift dependence of the merger rate 
using this approach.  

\section{PROPERTIES OF CLOSE PHYSICAL PAIRS} \label{pp}

In the following section, we will focus on close pairs in which both galaxies 
are members of the primary sample.  In addition, we will concentrate only on 
those pairs in which the galaxies are physically 
associated (based on the relative velocity criterion outlined in \S~\ref{vcut}). 
These pairs are the best candidates for systems undergoing mergers or 
interactions.  Images of the 13 pairs satisfying these criteria are displayed in 
Figure~\ref{figim}.  

Close inspection of these images reveals clear evidence of ongoing 
interactions in some pairs.  
While the image quality is not sufficient to perform detailed morphological 
classifications of each galaxy, it is adequate to separate the pairs 
in the following three categories : 
interacting (int), possibly interacting (int?), 
or not interacting (no).  Pair classification was carried out by DRP, CJP, and 
HKCY, with no 
prior knowledge of any other properties of the pairs or member galaxies.  
In all cases, at least two of us agreed on the classification, and there was 
no disagreement by more than one class.  The majority classification was 
adopted in each case.  

The pair identification number, rest-frame velocity difference, 
projected separation, and image classification for each pair is 
listed in Table 4. 
In addition, the galaxy identification number, rest-frame color, 
spectral classification, and [\ion{O}{2}] rest-frame equivalent width 
is given for each galaxy.  
It is immediately apparent that the four pairs classified as interacting (int)
are very blue, with exceptionally strong emission lines.  In addition, 
all four pairs have low ($<$ 150 km/s) relative velocities.  Since each 
of these measures is determined independently, this suggests
that these pairs are in the process of merging. 
Such strong evidence of merging is surprising in the light of earlier analysis 
on the mean properties of this sample.  That is, while the mean properties of 
these galaxies are indistinguishable from the field, a subset
of the pairs do exhibit significantly different properties.  
This shows that indications of mergers are being swamped 
by non-merging pairs in our sample of close pairs.

We clearly have not identified all galaxies in the sample which are 
currently affected by mergers.  First of all, there are some merging pairs 
which would be missed by our approach.  Galaxies with very close ($< 1\arcsec$) 
companions may appear as isolated galaxies on our images.  
Also, some galaxies may have close companions which fall below the magnitude 
limit imposed in our study.  
In both cases, even if these companions are relatively small, 
it is possible that minor mergers may trigger strong starburst activity 
(Mihos \& Hernquist 1994\markcite{MH94}).  
Secondly, there are probably isolated galaxies in the field that have recently
undergone mergers, and no longer have detectable companions nearby.
These galaxies will obviously be missed in a study of close pairs.  
They may be identified as merger remnants if they have distorted appearances.  
However, as demonstrated by Mihos\markcite{M95} (1995), 
morphological signs of interactions 
are relatively short-lived, and can be difficult to detect at these redshifts. 
The presence of emission lines (such as [\ion{O}{2}]) can be used independently 
to detect on-going merger-induced starbursts, while other spectral 
indices (e.g. $H\delta$) 
serve as useful diagnostics of recent starburst activity (see Couch \& 
Sharples 1987\markcite{CS}).  Furthermore, AGB light resulting from a starburst
may last for several Gyr after a merger (Silva \& Bothun 1995\markcite{SB}).
Finally, we note that some mergers may not produce any obvious signs that a 
merger
is taking place.  This is particularly true for mergers of early-type galaxies, 
which will not produce either a strong starburst or clear morphological signs of 
interaction. 

\section{DISCUSSION} \label{discuss}
Using a redshift-selected sample of field galaxies with $<z>$ $\sim$ 0.33, 
we find that 
roughly 7 \% have close physical companions of comparable luminosity.  
This is a factor of 1.7 higher than the physical pair fraction found locally.  
The mean properties of these paired galaxies are found to be
indistinguishable from those of isolated galaxies.
Upon dividing the sample into two independent sets based on redshift, 
we find a similar result.
The only trend we find, although not statistically significant, is for paired 
galaxies 
to be (on average) slightly redder, more luminous, and of earlier 
type-classification than galaxies with no close companions.  
These results are consistent 
with evidence that early-type galaxies are more clustered than late-type 
galaxies (Roberts
\& Haynes 1994\markcite{RH}).  For example, 
Loveday {\it et al.}\markcite{LV} (1995) find that early-type galaxies are 
clustered more
strongly than late-type galaxies by a factor of 3.5 $-$ 5.5 locally, while 
Neuschaefer {\it et al.}\markcite{NEU} (1995) 
find a similar effect in a sample with an average redshift of $z \sim$ 0.5.

Upon examining the images of the 13 pairs which are known to be physically 
associated (based on similar redshifts), we find that the galaxies in pairs 
which 
are classified as interacting (based on their appearances) do exhibit very 
different properties.  These pairs all have low relative velocities, and their 
galaxies have very blue rest-frame colors and spectra with strong emission
lines.  This is the first clear evidence (at this redshift) that paired galaxies
have significantly different properties than field galaxies in general.  
This implies that mergers are responsible for at least some of the 
starbursting activity seen in samples of faint field galaxies.  
There are two ways to improve on this result.  First, a larger and more complete 
redshift sample will increase the yield of close physical pairs.  Secondly, high 
resolution images would allow for improved detection of signs of interactions in 
these pairs.  

We can use the information in Figure~\ref{fig2z}b to make a rough estimate of 
the mean dynamical friction timescale for the close physical pairs in our 
sample.  Following Binney \& Tremaine (1987\markcite{BT}), we can estimate 
the time needed for a companion to merge.  Taking projected separations and 
relative velocities from Table 4, and assuming conservative values for the 
Coulomb logarithm (ln$\Lambda \sim 3$) and galaxy mass-to-light ratio 
($M/L \sim 5$), we find a mean timescale of $\sim$ 150 Myr.  Strictly speaking, 
this technique doesn't apply to mergers between galaxies of comparable mass 
(the typical mass ratio is 2:1 for these pairs); however, it confirms the 
assumption that the timescales involved are short.
An alternative approach is to use the method outlined by Charlton and 
Salpeter (1991\markcite{CS91}) for mergers between galaxies of comparable 
mass.  Assuming the inward 
drift due to dynamical friction is independent of separation (once halos overlap 
appreciably), and adopting a typical number of decay orbits (3) based on N-body 
calculations, we find a mean dynamical friction timescale of $\sim$ 400 Myr.  
Again, while this method is approximate in nature, it indicates that these close 
physical pairs should merge quickly, as expected.

The merger rate is estimated to increase with redshift as $(1+z)^m$, 
with m = 2.8 $\pm$ 0.9.  
We demonstrate that our determination of the increase in the merger rate with 
redshift is consistent with earlier determinations using close pairs, 
and provides one of the strongest constraints 
to date.  The observed increase in the merger rate with redshift implies that 
the average galaxy mass at $z \sim$ 0.33 was substantially lower.
Our observed value of m is consistent with the theoretical value of m = 2.5, 
derived by Toomre (1977\markcite{T77}) using binding energies of bound pairs.  
Carlberg (1990\markcite{C90}) estimated that m $\simeq 4.51 \Omega^{0.42}$, 
for $\Omega$ near 1 and a CDM-like cosmology.  
This implies that our result is consistent with a low density universe, as found 
in several recent studies (e.g. Carlberg {\it et al.} \markcite{C96}1996).  
Our result is also consistent with merging being the cause of 
the observed evolution in co-moving luminosity density found 
by Lilly {\it et al.} (1996\markcite{L96}).  For blue wavelengths, they find 
{\it L} $\propto (1+z)^{2.3 \pm 0.5}$ for $\Omega_0$ = 0.1 and $q_0$ = 0.05 
({\it L} $\propto (1+z)^{2.7 \pm 0.5}$ for $\Omega_0$ = 1 and $q_0$ = 0.5).  
They attribute this evolution to a rapid decline in
the global star formation rate since $z \sim$ 1.  

There are several obvious ways of improving on the results of this paper.  
A larger and more complete set of redshifts would decrease the 
swamping effect of unrelated optical pairs, 
as well as simply reducing the uncertainty in 
estimating the merger rate and differences in various properties.  
Secondly, additional 
spectral features and colors may prove to be better diagnostics of 
on-going or recent mergers. 
Finally, a significant short-coming of this type of study is simply a 
lack of resolution.  
High resolution imaging would allow close faint companions to be 
identified, purifying the pair sample.  
In addition, improved morphological indications of ongoing 
mergers or disruptions would help to identify merger candidates.  
Quantitative determinations of morphology of 
individual galaxies would also allow the competing effects of mergers and 
the morphology-density relation to be disentangled more easily.   

\acknowledgments
We wish to thank other members of the CNOC consortium for assistance in 
carrying out the observations and data reduction.  
We are grateful to Mike Balogh and Luc Simard for their assistance in 
measuring [\ion{O}{2}] line strengths.   We also wish to thank Mike
Hudson, Ron Marzke, and Simon Morris for helpful discussions. 
This work was supported by the 
Natural Sciences and Engineering Research Council,
through research grants to CJP, HKCY and RGC, 
and a post-graduate scholarship to DRP.

\newpage

\centerline{FIGURE CAPTIONS}
\figcaption[figsl.eps]
{The selection function for all galaxies in the region of the cluster E1621.  
The lines represent the 
restrictions imposed to ensure at least 25 \% spectroscopic completeness 
for each galaxy in the primary sample. \label{figsl}}

\figcaption[fignd.eps]
{Histogram of the local number density of galaxies 
within a $1\arcmin$ radius of each
unbiased field galaxy, using a limiting magnitude of $r$ = 22.0. \label{fignd}}

\figcaption[figrs.eps]
{Histogram of the pair number density, binned in annular rings of 
width 0\arcsecpt 5.
The lines represents the pair number density predicted using the $F$ 
band angular correlation 
function from Infante \& Pritchet (1995), with $\delta$ = 0.8 
({\it solid line}) and $\delta$ = 0.6 ({\it dotted line}).   
Note the absence of pairs at very small angular separations, 
due to the limited resolution 
of the observations.  ($a$) All galaxies brighter than $r$ = 23.0.  ($b$) 
All galaxies brighter than $r$ = 23.0, 
excluding known cluster members. \label{figrs}}

\figcaption[fig2z.eps]
{Projected separation (D) versus rest-frame velocity difference ($\Delta V$) 
for galaxy pairs with 2 redshifts.  Points with small 
projected separations and small velocity differences represent pairs which
are physically associated.  $\Omega_0$ = 1 was assumed for this calculation. 
($a$) Pairs with two redshifts satisfying $\Delta V <$ 10000 km/s and 
D $<$ 100 h$^{-1}$ kpc.  ($b$) The subset of close physical pairs of 
primary galaxies.  \label{fig2z}}

\figcaption[figcc.eps]
{The determination of a color cutoff.  For a given primary galaxy at 
redshift $z$, all companions without redshifts which lie 
more than 0.15 magnitudes redward of the E/S0 sequence are assumed to lie at 
larger redshift.  Hence, any such companions are expected to be unrelated.  
The non-evolving E/S0 sequence ({\it solid line}) 
was determined by convolving the filters with a typical E/S0 galaxy 
spectrum from CWW.  
The dashed line lies 0.15 magnitudes redward of the E/S0 sequence.  
Filled symbols represent field galaxies from the
primary sample; open symbols represent field galaxies 
which are not part of the primary sample. \label{figcc}}

\figcaption[figzh.eps]
{Histograms of redshifts for the paired and isolated samples.
Paired galaxies have at least one companion of comparable luminosity. 
Paired (faint) galaxies have at least one companion brighter than $r$ = 23.0, 
but none of comparable luminosity. 
Isolated galaxies have no companions brighter than $r$ = 23.0. \label{figzh}}
 
\figcaption[figmc.eps]
{The color classification of galaxies in the primary sample.  Classification was
determined based on the observed $g-r$ colors and redshifts. 
The lines represent the colors 
of non-evolving galaxies derived by YEC by
convolving Gunn filter responses with the standard 
typical galaxy spectra 
of CWW.  The different galaxy sequences 
and corresponding classifications are: 
E ({\it solid line, open circles}), 
Sbc ({\it dotted line, filled circles}), 
Scd ({\it dashed line, open squares}), 
and Im ({\it long-dashed line, stars}). \label{figmc}}

\figcaption[fig0h.eps]
{Histograms of rest-frame $g-r$ colors for the paired and isolated samples. 
Rest-frame colors were derived using 
$g-r$ $k$-corrections from YEC (using spectra of CWW).
Paired galaxies have at least one companion of comparable luminosity. 
Paired (faint) galaxies have at least one companion brighter than $r$ = 23.0, 
but none of comparable luminosity. 
Isolated galaxies have no companions brighter than $r$ = 23.0. \label{fig0h}}

\figcaption[figmh.eps]
{Histograms of absolute $r$ magnitudes for the paired and isolated samples.
Determination of M$_r$ uses $r$ band $k$-corrections derived from Sebok 
(1986) by YEC, and assumes $h$ = 1.
Paired galaxies have at least one companion of comparable luminosity. 
Paired (faint) galaxies have at least one companion brighter than $r$ = 23.0, 
but none of comparable luminosity. 
Isolated galaxies have no companions brighter than $r$ = 23.0. \label{figmh}}

\figcaption[figeh.eps]
{Histograms of [OII] rest-frame equivalent widths for the paired and 
isolated samples.
Paired galaxies have at least one companion of comparable luminosity. 
Paired (faint) galaxies have at least one companion brighter than $r$ = 23.0, 
but none of comparable luminosity. 
Isolated galaxies have no companions brighter than $r$ = 23.0. \label{figeh}}

\figcaption[figim.eps]
{Mosaic of images for 12 close 
{\it physical} pairs in which both galaxies 
are primary field galaxies.  North is up and east is to the left.  
Each image is 40 pixels square, corresponding to $\sim$ 12\arcsecpt 5 on a side. 
Pair identifications correspond to those found in Table 4. 
The physical projected pair separation ranges from 4 h$^{-1}$ kpc (p06) 
to 19 h$^{-1}$ kpc (p11). 
The typical FWHM is $\sim$ 1\arcsecpt 0, and is fairly constant from 
image to image.
\label{figim}}

\end{document}